\documentclass{ragtime}

\usepackage{txfonts}
\usepackage{amsmath}
\usepackage{mathtools}
\usepackage{amssymb}
\usepackage{graphicx}
\usepackage{float}
\usepackage{color}
\usepackage[mathscr]{euscript}
\usepackage{times}
\usepackage{dirtytalk}
\usepackage[T1]{fontenc}
\usepackage[utf8]{inputenc}
\usepackage[polish,english]{babel}
\hypersetup{
        colorlinks=true,
        allcolors=blue,
        }

\newcommand{\beq}{\begin{equation}}
\newcommand{\eeq}{\end{equation}}
\newcommand{\ba}{\begin{eqnarray}}
\newcommand{\ea}{\end{eqnarray}}
\newcommand{\alp}{\alpha}
\newcommand{\eps}{\epsilon}

\definecolor{DarkViolet}{RGB}{148,0,211}
\newcommand{\Color}{DarkViolet}
\hypersetup{colorlinks,bookmarksopen,bookmarksnumbered,citecolor={\Color},linkcolor={\Color},urlcolor={\Color},pdfstartview=FitH}

\title[Theory-agnostic parametrization of wormhole spacetimes]{Theory-agnostic parametrization of wormhole spacetimes}

\author[T.~D.~Pappas]{Thomas~D.~Pappas\at[]{}\\
\ins{}Research Centre for Theoretical Physics and Astrophysics, Institute of Physics,\splitins[1]  Silesian University in Opava, Bezru\v{c}ovo n\'am\v{e}st\'i 13, CZ-746 01 Opava, Czech Republic\\
\ins{}\Email{thomas.pappas@physics.slu.cz}}

\coentry{T.~Pappas}{Theory-agnostic parametrization of wormhole spacetimes}

\begin{document}

\begin{abstract}
We present\footnote{Talk given by the author at RAGtime 23, Opava, Czech Republic, based on~\cite{Bronnikov:2021liv}.} a generalization of the Rezzolla-Zhidenko theory-agnostic parametrization of black-hole spacetimes to accommodate spherically-symmetric Lorentzian, traversable wormholes (WHs) in an arbitrary metric theory of gravity. By applying our parametrization to various known WH metrics and performing calculations involving shadows and quasinormal modes, we show that only a few parameters are important for finding potentially observable quantities in a WH spacetime. 
\end{abstract}



\section{Introduction \label{sec:intro}}

Wormholes (WHs) are hypothetical tunnel-like spacetime structures that connect two different regions of our Universe and can even be envisaged as bridges between different universes~\cite{Visser:1995cc}. The concept of a WH emerged as early as in 1916 in the work by~\cite{Flamm:1916}, and also in the works of~\cite{Einstein:1935tc} and~\cite{Wheeler:1955zz}, however, these WHs were non-traversable. The first examples of exact solutions in GR corresponding to traversable WHs sourced by a phantom scalar field have been obtained in~\cite{Bronnikov:1973fh,Ellis:1973yv}. A significant rekindling of interest on the subject came about with the work of~\cite{Morris:1988cz}.

As the work of Morris and Thorne, and subsequent studies have revealed, traversable WHs typically come with a number of problems such as the requirement of exotic forms of matter to  support the throat from collapsing~\cite{Morris:1988cz}, and/or dynamical instabilities~\cite{Bronnikov:2002sf,Gonzalez:2008wd,Bronnikov:2011if,Bronnikov:2012ch,Bronnikov:2018vbs,Cuyubamba:2018jdl}. To date, \emph{there is no known fully satisfactory model for a traversable, Lorentzian WH that has been proven to be both free of the necessity for exotic mater for its existence and at the same time corresponding to a dynamically stable configuration}. As a consequence, a general theory-agnostic approach to parametrizing a WH geometry in such a way that can be constrained by current and upcoming observations could provide a solution. For a recent review on the search for astrophysical WHs in our Universe, the reader is directed to~\cite{Bambi:2021qfo}.

In the case of black hole (BH) geometries, a general theory-agnostic parametrization that can be constrained from observations has been proposed by~\cite{Rezzolla:2014mua} (RZ). This method is similar in spirit with the parametrized post-Newtonian formalism albeit with validity  that is not limited  to the weak-field regime but rather covers the whole spacetime outside the event horizon of the BH. The RZ BH metric is defined in terms of a compact radial coordinate and a continued-fraction expansion involving an infinite tower of dimensionless parameters. Due to the rapid convergence properties of the continued-fractions however, in practice, only the first few of the expansion parameters are dominant and important for describing observable quantities in a BH background. Inspired by the above, in~\cite{Bronnikov:2021liv}, we proposed a modification of the RZ BH metric that allows for the parametrization of WH geometries in a theory-agnostic way.

This article is structured as follows. In Sec.~\ref{Sec:II} we discuss in general terms some features of asymptotically-flat WH spacetimes. In Sec.~\ref{Sec:III}, we provide a brief overview of the RZ BH parametrization and then introduce our extended parametrization for WH. In Sec.~\ref{Sec:IV} we obtain parmetrizations for examples of known WH metrics. Section~\ref{Sec:V} is dedicated to the study of shadows and quasinormal modes on the parametrized WH backgrounds as tests for the accuracy of our method. We conclude in Sec.~\ref{Sec:Conclusions}.

\section{Wormhole spacetimes: General considerations}\label{Sec:II}

\subsection{Asymptotically flat, traversable Lorentzian wormholes}

The line element for an arbitrary four-dimensional static, spherically symmetric geometry can be written as
\beq
ds^2=-f(r) dt^2 +\frac{1}{h(r)} dr^2 +K^2(r) \left( d\theta^2+ \sin^2 \theta\, d\phi^2 \right) \,.
\label{gen_l_e}
\eeq
Out of the three metric functions $f(r)$, $h(r)$, $K(r)$ in the above ansatz, only two are independent, and upon appropriately transforming the radial coordinate, any metric can be cast in the form where $K(r)=r$, however this might not always be feasible analytically. In general, the area of the sphere at radial coordinate $r$ is $A(r)=4\pi\, K^2(r)$. A WH structure is characterized by a minimum radius $r_0$ called the \emph{throat} (narrowest part of the tunnel) for which the surface area is minimized, namely
\beq
A'(r_0)=0 \rightarrow K'(r_0)=0\,.
\eeq
The metric functions $f(r)$ and $h(r)$ are regular and positive in a range of $r$ containing the throat and values of $r$ on both sides from the throat such that $K(r) \gg K(r_0)$. It is then said that the metric \eqref{gen_l_e} describes a \emph{traversable, Lorentzian WH}. Furthermore, a WH is classified as being asymptotically flat if, for $r$ tending to some $r = r_\infty$, the following conditions are satisfied
\beq             
f(r) \rightarrow 1\,, \qquad h(r) \left(\frac{dK(r)}{dr} \right)^2 \rightarrow 1\,.
\label{asflat}
\eeq

\subsection{The Morris-Thorne frame}

A very commonly used frame in the literature where WH metrics are written is the one introduced by~\cite{Morris:1988cz} (MT)
\beq
ds^2=-e^{2\Phi(r)}dt^2+\left(1-\frac{b(r)}{r} \right)^{-1} dr^2 +r^2 d\Omega^2\,.
\label{MT_ansatz_le}
\eeq
There are two arbitrary metric functions in the above line element. The first, $\Phi(r)$, is often called the \emph{redshift function}, and absence of event horizons (WH traversability) requires that it should be finite everywhere. The second, $b(r)$, is called \emph{shape function}, and indirectly determines the spatial shape of the WH in its embedding diagram representation. In the MT frame, $r_0$ is determined by the condition
\beq
h(r) = \left(1-\frac{b(r)}{r} \right)=0\,,
\label{MT_r0}
\eeq
while $e^{\Phi(r_0)} > 0$ is required, and the radial coordinate is defined for $r \in [r_0,\infty)$. The shape function should satisfy the so-called flair-out conditions on the throat, $b(r_0)=r_0$ and $b'(r_0)<1$ while $b(r)<r$ for $r \neq r_0$. In the framework of GR,~\cite{Morris:1988cz} showed that WHs require the presence of some sort of exotic matter that violates the null energy condition. For recent developments regarding traversable WHs without exotic matter in Einstein-Dirac-Maxwell theory see~\cite{Blazquez-Salcedo:2020czn,Bolokhov:2021fil,Konoplya:2021hsm}.

\subsection{Wormhole shadows}

In this section, we outline the method for the computation of shadows in an arbitrary static spherically symmetric and asymptotically flat spacetime~\cite{Synge:1966okc,Perlick:2015vta}. Starting with the general metric ansatz~\eqref{gen_l_e}, it is convenient to introduce the function
\beq
w^2(r) \equiv \frac{K^2(r)}{f(r)}\,.
\label{w^2}
\eeq
The photon-sphere radius $r_{\rm ph}$, corresponds to the minimum of $w^2(r)$ and is thus determined as a solution to
\beq
\frac{d w^2(r_{\rm ph})}{dr}=0\,.
\label{uph_condition}
\eeq
The angular radius of the shadow (associated with the outermost photon sphere), as seen by a distant static observer located at 
$r_{\rm O}$, is then obtained by means of~\eqref{w^2} as
\beq
\sin^2{a_{\rm sh}} = \frac{w^2(r_{\rm ph})}{w^2(r_{\rm O})}\,.
\eeq
Under the assumption $r_{ \rm O} \gg r_0$, where $r_0$ is a characteristic length scale that can be identified with the radius of the WH throat, or the BH event horizon depending on the nature of the compact object under consideration, we have that
\beq
f(r_{\rm O}) \simeq 1\,,\quad K^2(r_{\rm O}) \simeq r_O^2\,,
\eeq
and thus one finds that the radius of the shadow is given by
\beq
R_{\rm sh} \simeq r_{\rm O} \sin{a_{\rm sh}} 
\simeq w(r_{\rm ph})= \frac{K(r_{\rm ph})}{\sqrt{f(r_{\rm ph})}}\,.
\label{Rsh}
\eeq

\section{Continued-fraction parametrization for wormholes}\label{Sec:III}

\subsection{The Rezzolla-Zhidenko parametrized black-hole metric}

Let us begin this section, by briefly reviewing the parametrization of spherically symmetric BHs suggested in \cite{Rezzolla:2014mua} (RZ), and subsequently we will see which modifications of this approach are required when going over to WH geometries. The RZ parametrization is based on a dimensionless compact coordinate (DCC) that maps $[r_0,\infty) \rightarrow [0,1]$ according to
\beq
x(r)\equiv 1-\frac{r_0}{r}\,,
\label{RZ_DCC}
\eeq
where $r_0$ is the location of the outer event horizon of the BH determined via the condition $f(r_0)=0$. If $K^2(r)=r^2$, then $r_0$ is also the radius of the outer event horizon. In terms of~\eqref{RZ_DCC}, the following parametrization equations are introduced:
\ba
f(r)&=&\widetilde{A}(x)\,,\label{RZ_param_eq_A}
\\
\frac{1}{h(r)}&=&\frac{\widetilde{B}(x)}{\widetilde{A}(x)}\,,
\label{RZ_param_eq_B}
\ea
where the parametrization functions $\widetilde{A}(x)$ and $\widetilde{B}(x)$ are defined as
\ba
\widetilde{A}(x) &\equiv& x \left[ 1-\eps (1-x)+(a_0-\eps)(1-x)^2
+\frac{a_1 }{1+\frac{a_2 x}{1+\frac{a_3 x}{\ldots}}}(1-x)^3  \right]\,,      \label{RZ_param_fun_2}
\\
\widetilde{B}(x) &\equiv&\left[ 1+b_0 (1-x) 
+ \frac{b_1 }{1+\frac{b_2 x}{1+\frac{b_3 x}{\ldots}}}(1-x)^2 \right] ^2\,.   \label{RZ_param_fun_3}
\ea
The above parametrization involves two families of parameters. The first family consists of three ``asymptotic'' parameters $(\eps, a_0,b_0)$, which are determined via the expansions of the parametrization equations at spatial infinity ($x=1$), while the second family consists of the remaining parameters $(a_1,a_2,\ldots ,b_1,b_2,\ldots)$ i.e. the ``near-field'' parameters which are determined at the location of the event horizon ($x=0$). For the axially-symmetric generalization of the RZ metric see~\cite{Konoplya:2016jvv}, while for its higher-dimensional extension see~\cite{Konoplya:2020kqb}. More recently, an extension of the parametrization to non-asymptotically flat cases has been proposed in~\cite{Konoplya:2022kld,Konoplya:2023kem}.

\subsection{The parametrized wormhole metric}

To construct our WH parametrization, we consider the radial coordinate compactification according to~\eqref{RZ_DCC}, with $r_0$ interpreted in this context as the location of the WH radius\footnote{For alternative definitions of the DCC and its optimization see~\cite{Bronnikov:2021liv}.}. Then, we may parametrize the metric functions according to~\cite{Bronnikov:2021liv}

\beq
f(r)=f_0+x \left[ \left(1-f_0 \right)-\left(\eps +f_0 \right) (1-x)+(a_0-\eps-f_0)(1-x)^2+\frac{a_1 (1-x)^3}{1+\frac{a_2 x}{1+\frac{a_3 x}{\ldots}}} \right]\,,
\label{WH_param_f}
\eeq

\beq
h(r)=h_0+x\left[ \left(1 -h_0\right)-\left(b_0+h_0 \right) (1-x) + \frac{b_1 (1-x)^2 }{1+\frac{b_2 x}{1+\frac{b_3 x}{\ldots}}} \right] \,.
\label{WH_param_h}
\eeq

Being an extension of the RZ parametrization, it is no surprise that our parametrized WH metric shares several appealing properties with its BH predecessor, to which it reduces in the limit $(f_0,h_0) \to (0,0)$. Quite importantly, it is valid for all space ($x\in [0,1]$), not only near $x=0$ or $x=1$, and the continued-fraction expansions, endow the parametrization with quick converge properties~\footnote{These properties, allow for the parametrization to also be utilized for the analytic representation of numerical WH solutions along the lines of the analyses performed for BH spacetimes, see e.g.~\cite{Younsi:2016azx,Kokkotas:2017zwt,Hennigar:2018hza,Konoplya:2019goy,Konoplya:2019fpy}.}. The $n-$th order approximation of a given metric can be easily obtained by setting the $(n+1)-th$ near-field parameters $(a_{n+1},b_{n+1})$ equal to zero, thus removing all the higher-order parameters from the expressions of the metric functions. The metric~\eqref{WH_param_f}-\eqref{WH_param_h} involves once again two families of parameters, the asymptotic ($\eps, a_0, b_0$), which are determined at ($x=1$) and the set ($f_0, h_0, a_i, b_i$) which are determined near the throat of the WH ($x=0$), in analogy to the BH case discussed in the previous section.

\subsection{Observational constraints on the asymptotic parameters}
Given that the parametrization is developed in a theory-agnostic way, i.e. independently of the underlying theory of gravity, there are no precise constraints to be imposed on the metric functions $f(r)$ and $h(r)$. However, general constrains on the asymptotic parameters can be imposed, via the parameterized post-Newtonian (PPN) expansions~\cite{Will:2005va,Will:2014kxa}. To this end, consider the expansions of our parametrized metric~\eqref{WH_param_f}-\eqref{WH_param_h} at $x=1$
\ba
f(r) &=& 1-\left(1+\eps \right) \left(1-x \right)+a_0 \left(1-x \right)^2+\mathcal{O}\left( \left(1-x\right)^3 \right)\,,\label{A(x)_WH_inf_exp}\\
\frac{1}{h(r)} &=& 1+\left(1+b_0 \right)\left( 1-x\right)+\mathcal{O}\left( \left(1-x \right)^2 \right)\,.\label{B(x)_WH_inf_exp}
\ea
It is then straightforward, by comparison with the PPN expansions, to associate the asymptotic parameters $(\eps,a_0,b_0)$ with the PPN parameters $\beta$ and $\gamma$ in the following way
\beq
\eps =\frac{2 M}{r_0}-1\,,\quad a_0=\frac{2M^2}{r_0^2} \left( \beta-\gamma \right)\,,
\eeq
and
\beq
b_0= \gamma \frac{2M}{r_0}-1\,\quad \Rightarrow \quad b_0= \gamma \left( \eps+1 \right)-1\,.    \label{PPN_b0}
\eeq
Since $\beta$ and $\gamma$ are constrained as $\left| \beta -1 \right| \lesssim 2.3 \times 10^{-4}$, and $\left| \gamma -1 \right| \lesssim 2.3 \times 10^{-5}\,$, it follows that in our parametrization, astrophysically viable WHs must be characterized by \textbf{$a_0 \simeq 0$} and \textbf{$b_0 \simeq \eps$}. This is to be contrasted with the PPN constraints on the BH parametrization for which one finds $a_0 \simeq 0$ and $b_0 \simeq 0$~\cite{Rezzolla:2014mua,Bronnikov:2021liv}.

\section{Examples of parametrization}\label{Sec:IV}

In this section, we consider various exact WH geometries and obtain the parametrizations for the first few orders in the continued-fraction expansion. This will provide a means to test the adequacy of the proposed method in providing accurate approximations for WH geometries in terms of only a few coefficients of the expansion. For more details and examples of WH parametrizations the interested reader is referred to~\cite{Bronnikov:2021liv}.

\subsection{The Bronnikov-Kim II braneworld wormhole solution}
In the context of the so-called Randall-Sundrum II braneworld model~\cite{Randall:1999vf}, by solving the Shiromizu-Maeda-Sasaki modified Einstein equations on the brane~\cite{Shiromizu:1999wj}, Bronnikov and Kim in~\cite{Bronnikov:2002rn} have obtained a large class of static, spherically symmetric Lorentzian wormhole solutions. Here we consider one of those solutions corresponding to a two-parametric family of spacetimes, with a line-element given by
\ba
ds^2&=&-\left(1-\frac{\alp^2}{r^2} \right)dt^2+\left(1-\frac{\alp^2}{r^2} \right)^{-1} \left(1+\frac{C-\alp}{\sqrt{2 r^2-\alp^2}} \right)^{-1} dr^2+r^2 d\Omega^2\,,\nonumber\\
&=&-\left(1-\frac{\alp^2}{r^2} \right)dt^2+\left(1-\frac{\alp^2}{r^2} \right)^{-1} \left(1-\sqrt{\frac{2 r_0^2-\alp^2}{2 r^2-\alp^2}}\right)^{-1} dr^2+r^2 d\Omega^2\,,
\label{BKzeromass}
\ea
where in the second line we have used the condition~\eqref{MT_r0} for the determination of the location of the WH throat $r_0$, in order to write $C=\alp-\sqrt{2 r_0^2-\alp^2}$. The above line element, is one with a zero Schwarzschild mass and exhibits black-hole and wormhole branches. For the WH branch, the absence of horizons implies $f(r) \equiv - g_{tt}(r)>0\, \forall \, r \in [r_0,\infty)$ and so the following condition between the parameters is established
\beq
f(r_0)=\left(1-\frac{\alp^2}{r_0^2} \right) \geqslant 0  \Rightarrow r_0 \geqslant \alp\,.
\eeq
The threshold between the WH and BH branches of the solution corresponds to $\alp=r_0$, where in this case, $r_0$ is identified with the location of the (double) BH event horizon. The WH/BH threshold is of special importance for testing the accuracy of the parametrization in the case of WHs that deviate only slightly from a BH geometry, thus corresponding to BH mimickers, see e.g.~\cite{Damour:2007ap,Churilova:2019cyt,Bronnikov:2019sbx}. For the metric function $f(r)$ the parametrization is exact\footnote{Note that, whenever a metric function has a polynomial form, the parametrization is, by construction, always exact at a finite order. This holds true for both the original RZ~\eqref{RZ_param_fun_2}-\eqref{RZ_param_fun_3} and our~\eqref{WH_param_f}-\eqref{WH_param_h} parametrized metrics.} with the values of the expansion parameters (EPs) being
\beq
\eps=-1\,,\quad a_0=-\frac{\alp^2}{r_0^2} \,,\quad f_0=1-\frac{\alp^2}{r_0^2}\,,\quad a_i =0\quad \forall i \geqslant 1\,.
\eeq
On the other hand, the parametrization of $h(r)$ is not exact, the first few EPs are
\ba
b_0&=&\sqrt{1-\frac{\alp^2}{2 r_0^2}}-1 \,,\quad h_0=0\,,\quad b_1 =b_0+ \frac{\alp^2}{\alp^2+2r_0^2}\,,\\
b_2 &=& \frac{2 \alp^4 \left(1+b_0 \right)-\alp^2 \left(7+8 b_0 \right) r_0^2+8 b_0 r_0^4}{b_1 \left( \alp^2 -2\, r_0^2 \right)^2}-3\,.
\ea

According to results presented in~Table~\ref{table:BKzeromass}, the first-order approximation, provides a very accurate description of the metric~\eqref{BKzeromass} with an absolute relative error (ARE) less than $1\%$ for the majority of the parametric space, i.e. for $p \lesssim 0.6$, but becomes less accurate as the WH/BH threshold is approached ($p \to 1$). However, it is also evident that the parametrization converges very quickly and as a consequence, the error is significantly reduced once the second-order correction is taken into account even at the WH/BH threshold~\cite{Bronnikov:2021liv}.
\begin{table}[H]
\center{
\caption{The maximum ARE in percents between the exact metric $h(r)$ and its approximation at various orders in terms of the dimensionless parameter $p \equiv \alp/r_0 \leqslant 1$. The WH/BH threshold corresponds to $p=1$.}
\begin{tabular}{c c c c c c c}
order & $p=0.1$ & $p=0.3$ & $p=0.5$ & $p=0.7$ & $p=0.8$ & $p=0.99$\\
1 & 0.00063 & 0.05460 & 0.49509 & 2.53408 & 5.39226 & 31.12707\\
2 & 0.00010 & 0.00840 & 0.07270 & 0.34042 & 0.67203 & 2.49612\\
3 & 0.00001 & 0.00118 & 0.00829 & 0.02370 & 0.02622 & 0.13332\\
4 &  $\mathcal{O}(10^{-8})$ & 0.00004 & 0.00093 &  0.00883 &  0.02312 &  0.12947\\
\end{tabular}
\label{table:BKzeromass}
}
\end{table}

\subsection{The Simpson-Visser geometry}

\begin{figure}[H]
\centering
\includegraphics[width=0.55\linewidth]{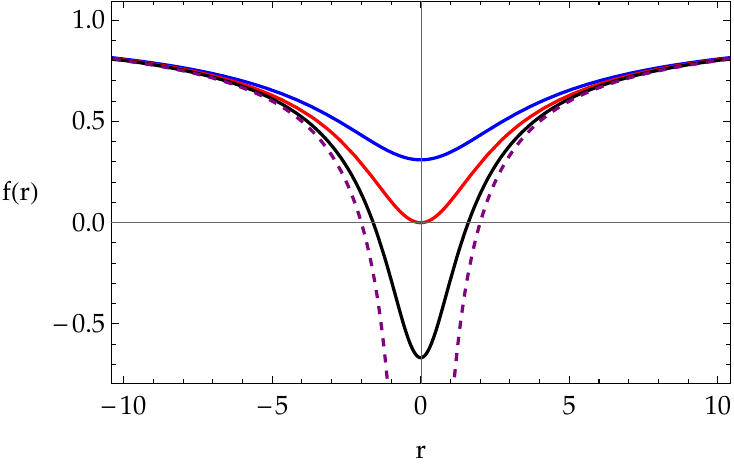}
\caption{The $f(r)$ metric function of the Simpson-Visser geometry~\eqref{SV_met_Orig} corresponding to a traversable wormhole (blue curve, for $p \equiv \alp/m=2.9$), an extremal regular BH (red curve, for $p=2$), a black bounce (black curve, for $p=1.2$), the Schwarzschild BH limit (dashed curve, for $p\to 0$).}
\label{fig:SV_branches}
\end{figure}

In~\cite{Simpson:2018tsi} (SV), an interesting geometry has been introduced as a toy-model via a one-parameter deformation of the Schwarzschild metric. Written in terms of the quasiglobal coordinate, the SV line element reads
\vspace{-0.3cm}
\beq
ds^2=-\left( 1-\frac{2\,m}{\sqrt{r^2+\alp^2}}\right) dt^2+\frac{dr^2}{\left( 1-\frac{2\,m}{\sqrt{r^2+\alp^2}}\right)} +\left(r^2+\alp^2\right) d\Omega^2\,.
\label{SV_met_Orig}
\eeq
The above line-element has been generalized to axial symmetry by~\cite{Mazza:2021rgq}, while the field sources for the SV metric have been obtained recently by~\cite{Bronnikov:2021uta}. Depending on the value of the dimensionless parameter $p\equiv \alp/m$, the SV metric describes a traversable WH for $p>2$, an extremal regular BH for $p=2$ (thus this value of $p$ defines also the WH/BH threshold), for $p<2$ a black-bounce state is obtained (see~\cite{Simpson:2018tsi}~\footnote{See also~\cite{Bronnikov:2005gm,Bronnikov:2006fu,Bolokhov:2012kn}.}), while for $p=0$ the Schwarzschild geometry is recovered, see also Fig.~\ref{fig:SV_branches}.

Notice that the metric~\eqref{SV_met_Orig} is not in the Morris-Thorne frame~\eqref{MT_ansatz_le} since $K^2(r)\neq r^2$. To this end, one may perform the coordinate transformation $r \rightarrow \tilde{r} : \tilde{r}^2 = r^2+\alp^2\,,$ in order to recast the metric to the MT frame where it is written as
\beq
ds^2=-\left( 1-\frac{2\,m}{\tilde{r}}\right) dt^2+\left( 1-\frac{2\,m}{\tilde{r}}\right)^{-1} \left( 1-\frac{\alp^2}{\tilde{r}^2}\right)^{-1} d\tilde{r}^2+\tilde{r}^2 d\Omega^2\,.
\label{SV_met_MT}
\eeq
Then, one may proceed with the parametrization in terms of~\eqref{RZ_DCC} and~\eqref{WH_param_h}. The condition~\eqref{MT_r0}, that determines the location of the WH throat $r_0$ in the MT frame, yields the following equation in the case of the SV wormhole
\beq
h(r) \equiv \left( 1-\frac{2\,m}{\tilde{r}}\right) \left( 1-\frac{\alp^2}{\tilde{r}^2}\right)=0\,.
\label{SV_MT_h(r)}
\eeq
There are two roots to the above equation, which are located at $\tilde{r}=2m$ and $\tilde{r}=\alp$. The former root, is not a suitable choice for a WH throat because in this case $f(\tilde{r}_0)=0$, and an event horizon emerges. Thus, the WH branch corresponds to the region in the parametric space defined by $\alp> 2m$. Since the metric functions in~\eqref{SV_met_MT} are of polynomial form, the parametrization is exact with the EPs given by
\ba
\eps&=&\frac{2m}{\alp}-1\,,\quad a_0=0 \,,\quad f_0=1-\frac{2m}{\alp}\,,\quad a_i =0\quad \forall i \geqslant 1\,,\nonumber\\
b_0&=&\frac{2m}{\alp}-1 \,,\quad h_0=0\,,\quad b_1=-\frac{2m}{\alp}\,,\quad b_i =0\quad \forall i \geqslant 2\,.
\ea
A general parametrization for WH metrics in non-MT frames by means of~\eqref{WH_param_f} and~\eqref{WH_param_h}, is also possible upon appropriate modification of the DCC. In particular, for the line-element~\eqref{SV_met_Orig}, the optimized version of the DCC \footnote{For more details on the DCC optimization see~\cite{Bronnikov:2021liv}.}
\beq
x(r)=1-\sqrt{\frac{R_0^2}{R_0^2+r^2}}\,,\qquad R_0=\alp \sqrt{\frac{3}{2}}\,,
\label{SV_ODCC}
\eeq
yields a parametrization for the SV metric with the first few EPs given by
\ba
\eps&=&\frac{2m}{R_0}-1 \,,\quad a_0=0  \,,\quad f_0= 1-\frac{2m}{\alp} \,,\quad a_1 = 2 \eps+3 f_0 -1+\frac{2 m R_0^2}{\alp^3}\,,\nonumber\\
a_2 &=&\frac{3}{2 a_1}\left( 4 \eps +5 f_0 -3 a_1 -1 +\frac{2m R_0^4}{\alp^5}\right)\,.
\label{SV_nonMT_EPs}
\ea

\section{Shadows and perturbations of test fields}\label{Sec:V}

In this section, as gauge-invariant tests for the accuracy of the parametrization, we consider shadows and perturbations of test fields in the background of the approximate WH metrics obtained by considering various orders in the continued-fraction expansion and compare them with the corresponding values obtained when the exact metric expressions are used.

\subsection{Shadows of the anti-Fisher wormhole}

In the context of GR with a massless minimally-coupled scalar field, a solution containing a naked singularity has been found~\cite{Fisher:1948yn}. When the kinetic term of the scalar field has the opposite sign, a solution emerges which has a WH branch and has been called anti-Fisher solution~\cite{Bronnikov:1973fh}. The metric functions for the latter solution assume the following form
\beq
f(r)=h(r)=e^{2u(r)}\,,\quad K^2(r)=e^{-2u(r)} \left(r^2+\alp^2 \right)\,,\quad u(r) \equiv \frac{m}{\alp}\left(\arctan\frac{r}{\alp}-\frac{\pi}{2} \right)\,.
\label{anti-Fisher_met_funcs}
\eeq
Substitution of~\eqref{anti-Fisher_met_funcs} in the general expression~\eqref{Rsh}, yields the shadow radius for the anti-Fisher WH
\beq
R_{\rm sh}= e^{-2\, u(r_{\rm ph})} \sqrt{4 m^2+\alp^2}\,,
\label{Rsh_anti-F}
\eeq
where $r_{\rm ph}=2m$ is the photon sphere radius which has been determined via~\eqref{uph_condition}. The first few EPs for the parametrization of $f(r)$ in this case are given by
\ba
\eps&=&1\,,\quad a_0=2 \,,\quad f_0=e^{2 u(m)}\,,\quad a_1 =f_0 \left( 3+\frac{2 m^2}{\alp^2+m^2} \right)-1\,,\nonumber \\
a_2 &=& \frac{6f_0-4 a_1-2}{a_1}\,,\quad a_3 = -\frac{f_0 \left[ \frac{2 m^4}{3 \left(\alp^2+m^2 \right)^2} -10 \right]+ a_1 \left[10+a_2 \left(5+a_2 \right)\right]+4}{ a_1 a_2} \,.
\label{anti-Fisher_EPs}
\ea

By considering various orders in the approximation of $f(r)$ via~\eqref{WH_param_f} and~\eqref{anti-Fisher_EPs}, we once again compute the shadow radius by means of~\eqref{Rsh} and compare the result order-by-order with the exact value given in Eq.~\eqref{Rsh_anti-F}. Our findings~\cite{Bronnikov:2021liv} are presented in terms of the dimensionless parameter $p\equiv \alp/m$ in Table~\ref{table:anti-Fisher_Rsh}. The high accuracy of the approximation already at the first order and the quick convergence are evident.

\begin{table}[H]
\center{
\caption{The percentage of absolute relative error between the exact value of the shadow radius~\eqref{Rsh_anti-F} and its value as obtained via Eq.~\eqref{Rsh} for various approximation orders of the metric function $f(r)$.}
\begin{tabular}{c c c c c c c c}
order &  $p=0.01$  & $p=0.2$ & $p=0.3$ & $p=0.4$ & $p=0.5$ & $p=0.6$\\
1 & 1.04775 & 0.95530 & 0.83863 & 0.67406 & 0.46096 & 0.19918\\
2 & 0.02753 & 0.03339 & 0.04295 & 0.06093 & 0.09281 & 0.14633\\
3 & 0.00304 & 0.00387 & 0.00570 & 0.01061 & 0.02367 & 0.05883\\
4 & 0.00011 & 0.00019 & 0.00034 & 0.00063 & 0.00112 & 0.00186\\
\end{tabular}
\label{table:anti-Fisher_Rsh}
}
\end{table}

\subsection{Shadows of the Simpson-Visser wormhole}

As a second example for shadows in a wormhole background we consider the SV metric in the non-Morris-Thorne frame i.e.~\eqref{SV_met_Orig}, for which we obtain the exact expression for the shadow radius
\beq
R_{\rm sh}=3\sqrt{3}m\,.
\label{R_sh_SV}
\eeq
Notice that value of $R_{\rm sh}$ is independent of the parameter $\alp$ and it is identified with the shadow radius of the Schwarzschild BH, for detailed discussions see~\cite{Tsukamoto:2020bjm,Junior:2021atr}. Subsequently, by considering various orders for the approximate metric according to~\eqref{SV_nonMT_EPs}, we compute once again $R_{\rm sh}$ and compare it with the exact result~\eqref{R_sh_SV}. The range of values for the dimensionless parameter $p\equiv \alp/m$ that is relevant for the analysis here is $p\in[2,3]$ where the lower bound corresponds to the WH/BH threshold and the upper bound corresponds to the maximum value of $p$ for which the spacetime under consideration exhibits a photon sphere. Our findings~\cite{Bronnikov:2021liv} are displayed in Table~\ref{SV_ODCC_R0_ip_Rsh}.

\begin{table}[H]
\center{
\caption{The percentage of absolute relative error between the analytic value of the shadow and its value for various approximation orders of the metric.}
\begin{tabular}{c c c c c c c}
order &  $p=2.01$ & $p=2.1$ & $p=2.4$ & $p=2.5$ & $p=2.7$ & $p=2.99$\\
1 & 0.54727 & 0.47968 & 0.25291 & 0.18463 & 0.07285 & 0.00009\\
2 & 0.01973 & 0.01568 & 0.00544 & 0.00329 & 0.00076 & $ \mathcal{O}(10^{-8})$\\
3 & 0.00278 & 0.00200 & 0.00046 & 0.00023 &  0.00003 & $\mathcal{O}(10^{-11})$\\
4 & 0.00060 & 0.00039 & 0.00005 & 0.00002 & $\mathcal{O}(10^{-6})$ & $\mathcal{O}(10^{-13})$\\
\end{tabular}
\label{SV_ODCC_R0_ip_Rsh}}
\end{table}

We observe that already with the first-order approximation of the metric, the error is less than $1\%$ for all values of $p$, and it is monotonically decreasing from the WH/BH threshold all the way to the no-photon sphere limit. Furthermore, one can see the quick convergence of the series where the error is reduced by approximately one order of magnitude with each additional term in the expansion that is taken into account.

\subsection{Quasinormal modes}

Let us now consider the fundamental quasinormal modes (QNMs) of the electromagnetic field propagating in a WH background. QNMs are characteristic frequencies of a compact object which are independent of the initial conditions of perturbations and are completely determined by the parameters of the compact object under consideration~\cite{Kokkotas:1999bd,Berti:2009kk,Konoplya:2011qq}. The real part of a QNM represents a real oscillation frequency, while the imaginary part is proportional to the damping rate. For a non exhaustive list of works where QNMs in WHs backgrounds have been studied, see~\cite{Konoplya:2010kv,Bronnikov:2012ch,Taylor:2014vsa,Cuyubamba:2018jdl,Volkel:2018hwb,Aneesh:2018hlp,Konoplya:2018ala,Kim:2018ang,Roy:2019yrr,Churilova:2019qph,Jusufi:2020mmy,Biswas:2022wah}. An electromagnetic field obeys the general covariant Maxwell equations
\beq
\dfrac{1}{\sqrt{-g}} \partial_\mu\left(F_{\rho\sigma}g^{\rho\nu}g^{\sigma\mu}\sqrt{-g}\right)=0.
\label{genelm}
\eeq
Here $F_{\rho\sigma}=\partial_\rho A_\sigma - \partial_\sigma A_\rho$ and $A_{\mu}$ is a vector potential. For the spherically-symmetric spacetime~\eqref{gen_l_e}, one may introduce the ``tortoise coordinate'' $r_*$, in terms of the metric functions $f(r)$ and $h(r)$ as
\beq
dr_*=\frac{dr}{\sqrt{f(r) h(r)}}\,.
\label{tortoise}
\eeq
Then, after separation of variables, Eq.~\eqref{genelm} assumes the following wave-like form
\beq
\dfrac{d^2 \Psi}{dr_*^2}+\left(\omega^2-V(r)\right)\Psi=0\,,
\label{klein-Gordon}
\eeq
and the effective potential reads as
\beq
V_{\rm em}(r) = f(r)\frac{\ell(\ell+1)}{r^2}\,.
\label{scalarpotential}
\eeq
Even though the effective potential depends on the gravitational background only via the metric function $f(r)$, the QNMs will depend on both $f(r)$ and $h(r)$ implicitly via the tortoise coordinate~\eqref{tortoise}. The boundary conditions for finding QNMs in a WH background correspond to purely outgoing waves at both infinities $r_* \rightarrow \pm \infty\,$~\cite{Konoplya:2005et}. The values of QNMs obtained for the Bronnikov-Kim II and Simpson-Viser WHs both of which have been discussed in the previous sections are displayed in Tables~\ref{QNM_BKII} and~\ref{QNM_SV} respectively~\cite{Bronnikov:2021liv}.

\begin{table}[H]
\center{
\caption{Fundamental quasinormal modes ($\ell=1$, $n=0$)  of the electromagnetic field for the Bronnikov-Kim II wormhole.}
\begin{tabular}{c c c c}
order & $C=-0.01$  & $C=-0.3$ &  $C=-0.7$\\
exact & $0.64306 - 0.20529 i$ & $0.70322 - 0.03991 i$ & $0.73338 - 0.09961 i$\\
$1$st & $0.65899-0.19306 i$  & $0.70232-0.03741 i$ & $0.73304-0.09836 i$\\
$2$nd & $ 0.64093-0.20593 i$  & $0.70332-0.04009 i$ & $0.73345-0.09969 i$\\
\end{tabular}\label{QNM_BKII}
}
\end{table}

\begin{table}[H]
\center{
\caption{Fundamental quasinormal modes ($\ell=1$, $n=0$)  of the electromagnetic field for the Simpson-Visser wormhole ($m=1/2$).}
\begin{tabular}{c c c c}
order & $\alp = 1.01$  & $\alp=1.25$ &  $\alp = 1.4$\\
exact & $0.51125 - 0.13311 i$ & $0.55491 - 0.03299 i$ & $0.56858 - 0.05986 i$\\
$1$st & $0.50899 - 0.13211 i$  & $0.55394 - 0.03297 i$  & $0.56770 - 0.05982 i$\\
$2$nd & $0.51241 - 0.13405 i$  & $0.55520 - 0.03286 i$  & $0.56885 - 0.05969 i$\\
\end{tabular}\label{QNM_SV}
}
\end{table}

As it can be seen, the QNMs obtained with the first two orders in the continued-fraction expansion of the background metric approximate very accurately the values obtained in terms of the exact metric.

\section{Conclusions}\label{Sec:Conclusions}
Building upon the~\cite{Rezzolla:2014mua} theory-agnostic parametrization of BH spacetimes, we have introduced an extension that allows for general Lorentzian, traversable, static and asymptotically-flat WH metrics to be accommodated in this parameterized framework~\cite{Bronnikov:2021liv}. The parametrization is based on a continued-fraction expansion in terms of a compactified radial coordinate, exhibits rapid convergence properties, and is valid in all of spacetime.

We have obtained the parametrizations for various examples of known wormhole geometries and studied the shadows and perturbations of test fields in these gravitational backgrounds for different orders in the expansion. Quite importantly, by considering geometries that interpolate continuously between a BH and a WH, we have demonstrated that the parametrization is very accurate, (already at the first and in some cases second-order in the expansion), even at the WH/BH threshold and this is relevant for WHs that act as BH mimickers.

Our analysis demonstrates that when a WH metric changes relatively slowly in the radiation zone, the observable effects in the wormhole background depend only on a few parameters and the following approximation for the line-element is sufficient~\cite{Bronnikov:2021liv}\footnote{We have also extended the general parametrized metric~\eqref{final01}-\eqref{final03} to accommodate slowly-rotating WHs~\cite{Bronnikov:2021liv}.}
\ba
ds^2 &=&-f(r) dt^2 +\frac{1}{h(r)} dr^2+ r^2  d\Omega^2 \,,\label{final01}\\
f(r) &=&1-\frac{r_0 \left(1+\eps \right)}{r}+ \frac{r_0^3 \left(a_1+f_0+\eps \right)}{r^3}-\frac{r_0^4\, a_1}{r^4}\,,\\
h(r) &=&1-\frac{r_0 \left(1+\eps \right)}{r}+ \frac{r_0^2 \left(b_1+h_0+\eps \right)}{r^2}-\frac{r_0^3\, b_1}{r^3}\label{final03}\,.
\ea

\section*{Acknowledgments}
The author would like to thank the organizers of RAGtime~23 for giving him the opportunity to present this work, and acknowledges the support of the Research Centre for Theoretical Physics and Astrophysics of the Institute of Physics at the Silesian University in Opava.

\bibliography{pap}
\end{document}